\documentstyle[psfig]{mn}

\def\ga{\mathrel{\hbox{\rlap{\hbox{\lower4pt\hbox{$\sim$}}}\hbox{$>$}}}}
\def\fd{\hbox{$.\!\!^{\rm d}$}}

\title[The new AM CVn star in Hydra]
{The new AM CVn star in Hydra}
\author[Patrick A. Woudt and Brian Warner]
       {Patrick A. Woudt\thanks{E-mail: pwoudt@circinus.ast.uct.ac.za} 
        and Brian Warner\thanks{E-mail: warner@physci.uct.ac.za}\\
        Department of Astronomy, University of Cape Town, Private Bag,
        Rondebosch 7700, South Africa}
\date{11 June 2003}

\begin{document}

\maketitle

\begin{abstract}
High speed photometry of the new AM CVn star in Hya (previously known as
SN2003aw), spectroscopically identified by Chornock \& Filippenko, shows it
to have a superhump period of 2041.5 $\pm$ 0.3 s.
We find a range of brightness from V $\sim$ 16.5 to 20.3, 
presumably caused by variations in the rate of mass transfer. 
In the intermediate state the system cycles in brightness with a period
of $\sim$ 16 h and range $\ge 0.4$ mag. There are sidebands to the principal 
frequencies in the Fourier transform which have constant frequency difference
from the superhump harmonics.
\end{abstract}


\section{Introduction}

\begin{table*}
 \centering
  \caption{The AM CVn Stars}
  \begin{tabular}{@{}llrrl@{}}
 Object       & V (mag)     & $P_{orb}$ (s)  & $P_{sh}$ (s)   & References \\[5pt]
{\bf RX\,J0806}  & 21.1     & 321.25$^*$     &                & Israel et al.~(2002); Ramsay, Hakala \& Cropper (2002)\\
{\bf V407 Vul}   & 19.9     & 569.38$^*$     &                & Cropper et al.~(1998)\\
{\bf ES Cet}     & 16.9     & 620.26         &                & Warner \& Woudt (2002) \\
{\bf AM CVn}     & 14.1     & 1028.7         & 1051.2         & Solheim et al.~(1998); Skillman et al.~(1999)\\
{\bf HP Lib}     & 13.7     & 1102.7         & 1119.0         & O'Donoghue et al.~(1994); Patterson et al.~(2002)\\
{\bf CR Boo}     & 13.0 -- 18.0& 1471.3        & 1487         & Wood et al.~(1987); Patterson et al.~(1997)\\
{\bf KL Dra}     & 16.8 -- 20 & 1500        & 1530            & Wood et al.~(2002)\\
{\bf V803 Cen}   & 13.2 -- 17.4 & 1612.0    & 1618.3          & Patterson et al.~(2000)\\
{\bf CP Eri}     & 16.5 -- 19.7 & 1701.2    & 1715.9          & Abbott et al.~(1992)\\
{\bf `2003aw'}   & 16.5 -- 20.3 &           & 2041.5          & This paper\\
{\bf GP Com}     & 15.7 -- 16.0 & 2974      &                 & Nather, Robinson \& Stover (1981); Marsh, Horne \& Rosen (1991) \\
{\bf CE-315}     & 17.6         & 3906        &               & Ruiz et al.~(2001); Woudt \& Warner (2002)\\[5pt]
\end{tabular}
{\footnotesize 
\newline 
Notes: $^*$ These have not yet been definitively established as orbital periods; $P_{orb}$ is the orbital period; $P_{sh}$ is the superhump period.\hfill}
\label{tab1}
\end{table*}

The AM CVn stars are recognized to be cataclysmic variable stars (CVs)
in which helium is being transferred from a degenerate donor to a 
degenerate accretor (see Warner (1995a) for a review of these stars).
They parallel the behaviour of the hydrogen-rich CVs, having 
high-rate-of-mass-transfer ($\dot{M}$) stable discs (the nova-like variables), unstable
high $\dot{M}$ systems (nova-likes of the VY Scl type), intermediate 
$\dot{M}$ dwarf novae, and very low $\dot{M}$ systems perhaps permanently
in a low state.
All of the AM CVn stars are remarkable for their short orbital periods --
ranging from possibly 5 min up to 65 min. The orbital modulations of 
brightness are of very low amplitude -- the dominant period in the light
curve is usually that of a superhump. This is the result of the very small
mass ratios in these stars, which causes their accretion discs to become
elliptical (e.g., Warner 1995a).

Until recently, there were nine definite and two possible AM CVn stars known,
listed in Table 1 with a small selection of relevant references. The 
spectroscopic accreditation of a further member of this class, found in 
a search for supernovae, was announced by Chornock and Filippenko (2003)
who observed the spectrum of supposed supernova 2003aw on the night of
2003 February 28, and found it to have a blue continuum with superposed
broad and weak He\,I emission lines at nearly zero redshift. KL Dra was
similarly discovered in a supernova search (Jha et al.~1998). Alerted
by Chornock and Filippenko's announcement, we made photometric observations
of the new candidate, which we will refer to in this paper as `2003aw'.

\section{Photometric observations}

We observed `2003aw' with the 40-in and 74-in reflectors at the Sutherland site of the South African Astronomical
Observatory (SAAO), using the University of Cape Town (UCT) CCD high speed photometer (O'Donoghue 1995) and no interposed filter, i.e., in `white
light'. We calibrated our instrumental magnitudes by observations of a hot standard star. In addition, Retha Pretorius
obtained some nightly brightness measurements for us, using the UCT CCD photometer on the 30-in reflector at Sutherland.
The observing log of the high-speed (and snapshot) photometry of `2003aw' is given in Table 2.

\begin{figure}
\centerline{\hbox{\psfig{figure=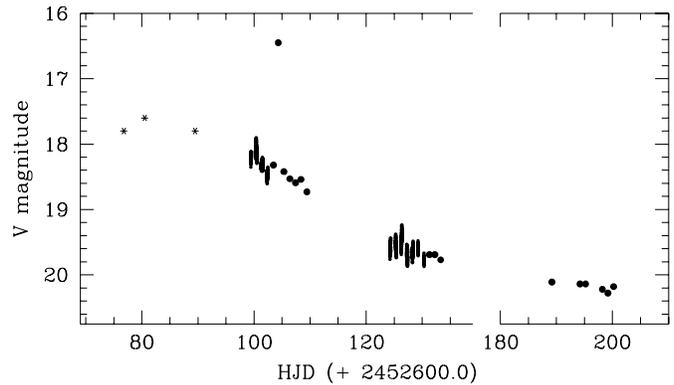,width=8.8cm}}}
  \caption{The long term light curve of `2003aw'.}
 \label{ltlc2003aw}
\end{figure}

\begin{figure*}
\centerline{\hbox{\psfig{figure=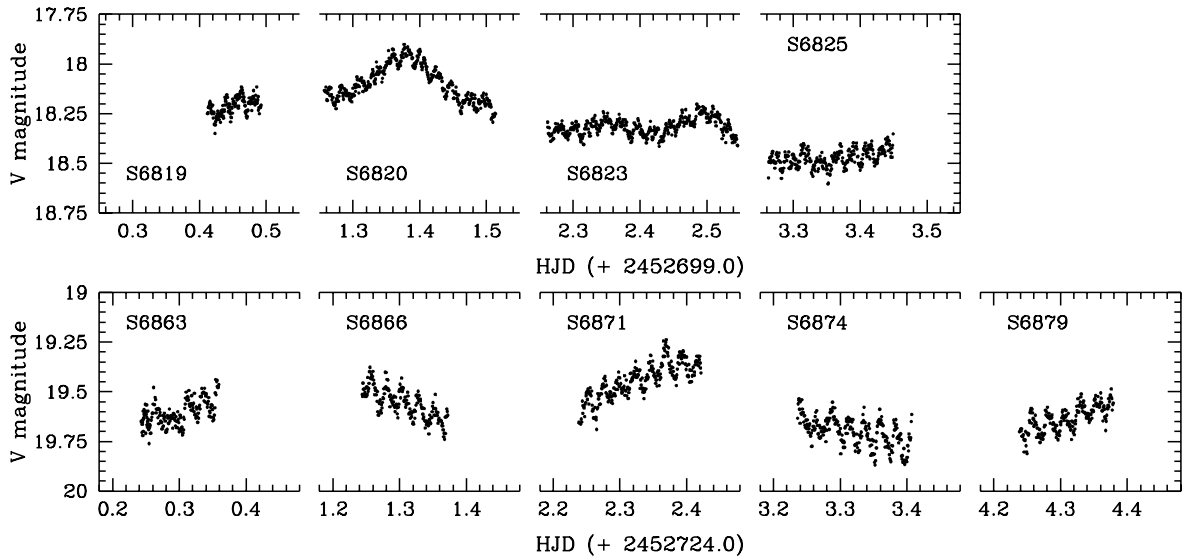,width=16.0cm}}}
  \caption{The light curves of `2003aw', obtained in 2003 February and March. The upper set of light curves were taken
when `2003aw' was still relatively bright, the lower set were taken when `2003aw' returned to a lower (intermediate) state.}
 \label{lc2003aw}
\end{figure*}

\section{Light curves}

\subsection{The longer-term brightness variations}

The long-term light curve, derived from the magnitudes listed in Table 2 and plotted in Fig.~\ref{ltlc2003aw}, shows
a range of $\sim$ 3.8 mag, but the length of coverage is too short to be sure that this represents the true full range.
Superimposed on the large slow changes of mean brightness are rises and falls of $\sim$ 0.3 mag during most of the
runs, i.e.,  on times scales of hours (see also Fig.~\ref{lc2003aw}). 
There is also a very bright state that occurred when the system was already in a fairly
high state, but lasted for less than a day. Apart from this last point, the light curve resembles those of V803 Cen (Patterson et al.~2000:
hereafter P2000) and CR Boo (Patterson et al.~1997: hereafter P1997), both of which show `high states' that bear some resemblance
to superoutbursts, long sessions of cycling at an intermediate level that have been said to resemble dwarf nova outburst behaviour, and
short visits to much lower states. In V803 Cen, the cycle time is $22 \pm 1$ h with a range of 1.1 mag, and in CR Boo the cycle time is
$\sim 19$ h, with a range of 1.1 mag. These properties were deduced from multi-site campaigns that were able to follow the cycling almost
continuously. From our single site observations we cannot be so definite about `2003aw', but with these two examples as our guide
we note that the variations within each night, and from night to night, in the intermediate state are well fitted with a cycle
time of $\sim 16$ h, but with an indeterminate range exceeding $\sim$ 0.4 mag.

\begin{table*}
 \centering
  \caption{Observing log of high speed and snapshot photometry of `2003aw'.}
  \begin{tabular}{@{}rrrrrcc@{}}
 Run No.  & Date of obs.          & HJD of first obs. & Length    & $t_{in}$ & Tel. &  V \\
          & (start of night)      &  (+2452000.0)     & (h)       &     (s)   &      & (mag) \\[10pt]
 S6819    & 2003 Feb 28 &  699.41240  &   2.19      &   45, 60  &  40-in & 18.2 \\
 S6820    & 2003 Mar 01 &  700.25748  &   6.15      &         60  &  40-in & 18.1  \\
 S6823    & 2003 Mar 02 &  701.26168  &   6.83      &         60  &  40-in & 18.3\\
 S6825    & 2003 Mar 03 &  702.26245  &   4.48      &         60  &  40-in & 18.5\\
 S6830    & 2003 Mar 04 &  703.46834  &             &        180  &  30-in & 18.3  \\
 S6834    & 2003 Mar 05 &  704.32556  &             &    60, 120  &  30-in & 16.5  \\
 S6839    & 2003 Mar 06 &  705.32117  &             &        120  &  30-in & 18.4  \\
 S6846    & 2003 Mar 07 &  706.39227  &             &   120, 300  &  30-in & 18.5  \\
 S6851    & 2003 Mar 08 &  707.41999  &             &   200, 300  &  30-in & 18.6  \\
 S6854    & 2003 Mar 09 &  708.35306  &             &        240  &  30-in & 18.5  \\
 S6861    & 2003 Mar 10 &  709.41747  &             &   180, 240  &  30-in & 18.7  \\
 S6863    & 2003 Mar 25 &  724.24288  &   2.79      &     30, 60  &  74-in & 19.6  \\
 S6866    & 2003 Mar 26 &  725.24377  &   3.08      &         60  &  74-in & 19.6  \\
 S6871    & 2003 Mar 27 &  726.23852  &   4.39      &         60  &  74-in & 19.4  \\
 S6874    & 2003 Mar 28 &  727.23629  &   4.58      &         60  &  74-in & 19.7  \\
 S6879    & 2003 Mar 29 &  728.23867  &   3.37      &         60  &  74-in & 19.6  \\
 S6884    & 2003 Mar 30 &  729.23656  &   0.86      &         60  &  74-in & 19.6  \\
 S6891    & 2003 Mar 31 &  730.31131  &   0.95      &         60  &  74-in & 19.8  \\
 S6894    & 2003 Apr 01 &  731.27721  &             &   120, 300  &  30-in & 19.7  \\
 S6899    & 2003 Apr 02 &  732.24219  &             &   120, 200  &  30-in & 19.7  \\
          & 2003 Apr 03 &  733.28889  &             &   180, 180  &  30-in & 19.8  \\
          & 2003 May 29 &  789.19547  &             &180, 200, 300&  30-in & 20.1  \\
          & 2003 Jun 03 &  794.19963  &             &         90  &  74-in & 20.1  \\
          & 2003 Jun 03 &  795.20332  &             &         90  &  74-in & 20.1  \\
          & 2003 Jun 07 &  798.19472  &             &         60  &  74-in & 20.2  \\
          & 2003 Jun 08 &  799.19673  &             &         60  &  74-in & 20.3  \\
          & 2003 Jun 09 &  800.19639  &             &         90  &  74-in & 20.2  \\
\end{tabular}
{\footnotesize 
\newline 
Note: $t_{in}$ is the integration time.\hfill}
\label{tab2}
\end{table*}

Although our observations are sparse, the evidence from Fig.~\ref{ltlc2003aw} is that `2003aw' had been in a low state
prior to its discovery as a possible supernova, brightened and remained at V $\sim$ 17.5 for about two weeks and steadily decreased
in brightness over the subsequent three months. In this it resembles VY Scl behaviour, in which $\dot{M}$ can change over a variety of
time scales.

\subsection{Rapid brightness variations}

Fig.~\ref{lc2003aw} shows the light curves obtained from our two weeks of high speed photometry, the first with the 40-in telescope
when the star had just been recognised as an AM CVn star (Chornock \& Filippenko 2003; Woudt \& Warner 2003) and was quite bright,
and the second when it had faded and we were fortunately observing with the larger telescope. 

\subsubsection{The superhump modulation}

In all of the light curves
there is a clear modulation with a period of 2041.5 s ($\pm$ 0.3 s) and range of $\sim$ 0.1 mag. This persistence of what we suppose to be
superhump modulation into lower states of mass transfer is similar to what is found in V803 Cen (P2000) and CR Boo (P1997).
The ephemeris for minimum light, derived from the data set obtained during the intermediate state, is given in Eq.~\ref{eq1}.

\begin{equation}
{\rm HJD_{min}} = 245\,2724.25281 + 0\fd023628 (\pm 3) \, {\rm E}.
\label{eq1}
\end{equation}

\begin{figure}
\centerline{\hbox{\psfig{figure=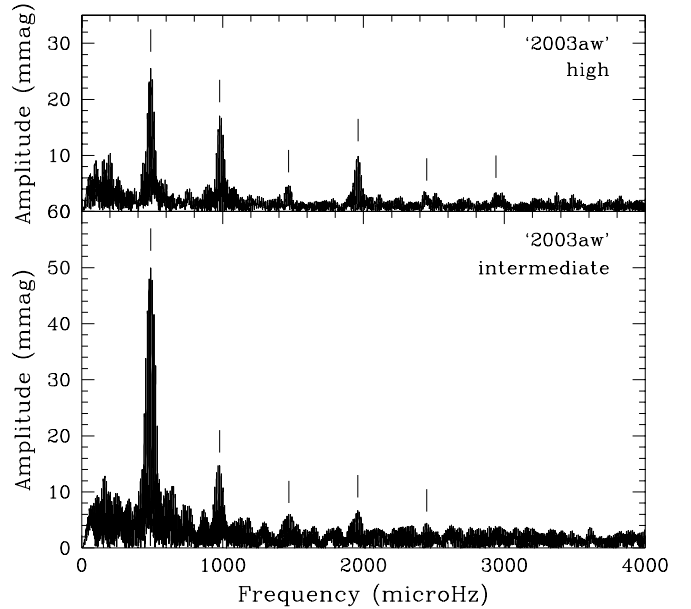,width=8.8cm}}}
  \caption{The Fourier transforms of `2003aw' in a high state (upper panel) and intermediate state (lower panel). The frequencies listed
in Table~\ref{tab3} are marked by vertical bars.}
 \label{2003awft}
\end{figure}

Fourier transforms (FTs) for the set of high state observations (runs S6819 -- S6825 of Table~\ref{tab2}) and for the intermediate
state observations (runs S6863 -- S6891) are shown in Fig.~\ref{2003awft}. For the light curves in the intermediate state, an individual mean
and slope has been subtracted in the FT. In the high state, runs S6820 and S6823 were prewhitened at the low frequency modulation visible in 
Fig.~\ref{lc2003aw}, after which individual mean values were subtracted in the FT. The fundamental and several harmonics show that the profile
of the superhump is highly non-sinusoidal, as has been found in other AM CVn stars. Table~\ref{tab3} lists the measured
frequencies (with uncertainties from non-linear least squares fits) and amplitudes of the fundamental and harmonics.

In our first observations, the dip near the maximum of the profile initially appeared to be drifting slowly relative
to the hump itself, which led us to announce the possibility of a shallow eclipse (Woudt \& Warner 2003). Later
observations, however, showed it to be a fixed feature in the profile, very similar to the profile in ES Cet 
(Warner \& Woudt 2002).

\begin{figure}
\centerline{\hbox{\psfig{figure=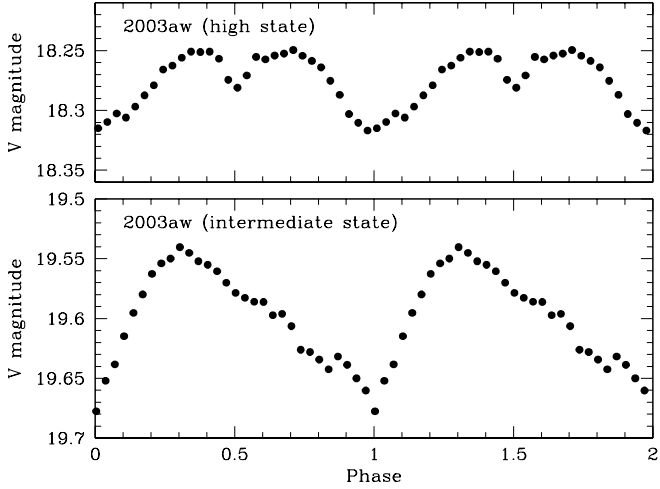,width=8.8cm}}}
  \caption{The average light curve of `2003aw' in a high state (upper panel) and intermediate state (lower panel). The latter is
phased on the ephemeris given in Eq.~\ref{eq1}; the high state light curve is relative to the same period, but phased independently
(due to the long gap between the two groups of observations) to match the minimum at intermediate state. }
 \label{2003awav}
\end{figure}

\begin{table}
 \centering
  \caption{Frequencies of the superhump modulation and its harmonics; amplitudes (in mmag) are given in square brackets.}
  \begin{tabular}{@{}ccc@{}}
                 & High state   & Intermediate state \\
ID               & Frequency    & Frequency  \\
                 & $\mu$Hz      & $\mu$Hz    \\[5pt]
\hfill {$\Omega_{sh}$}  & \hfill 489.91 $\pm$ 0.09 [25.1] & \hfill 489.84 $\pm$ 0.03 [49.9] \\
\hfill {2 $\Omega_{sh}$}& \hfill 979.54 $\pm$ 0.13 [16.9] & \hfill 979.70 $\pm$ 0.12 [14.7] \\
\hfill {3 $\Omega_{sh}$}& \hfill 1469.30 $\pm$ 0.46 [\, 4.6] & \hfill 1469.54 $\pm$ 0.28 [\, 6.0] \\
\hfill {4 $\Omega_{sh}$}& \hfill 1961.96 $\pm$ 0.22 [\, 9.8] & \hfill 1959.31 $\pm$ 0.24 [\, 7.1] \\
\hfill {5 $\Omega_{sh}$}& \hfill 2448.39 $\pm$ 0.75 [\, 2.8] &  \hfill 2449.34 $\pm$ 0.37 [\, 4.5] \\
\hfill {6 $\Omega_{sh}$}& \hfill 2940.40 $\pm$ 0.61 [\, 3.5] &  -- \\[5pt]

\end{tabular}
\label{tab3}
\end{table}

\subsubsection{Sidebands}

\begin{table}
  \caption{Sideband frequencies.}
  \begin{tabular}{@{}ccccc@{}}
                 & \multicolumn{2}{c}{High state}             & \multicolumn{2}{c}{Intermediate state} \\
ID               & Frequency  &   $\Delta \nu$ & Frequency    & $\Delta \nu$ \\
                 & $\mu$Hz    &   $\mu$Hz      & $\mu$Hz      & $\mu$Hz    \\[5pt]
\hfill {$\Omega_{sh} - \Delta \nu$}   & \hfill  470.5$\pm$0.4 [5.6] & 19.4      &  &\\
\hfill {3$\Omega_{sh} - \Delta \nu$}  & \hfill 1449.8$\pm$0.5 [4.4] & 19.5      & 1454.5$\pm$0.3 [4.9] & 15.0 \\
\hfill {5$\Omega_{sh} - \Delta \nu$}  & \hfill 2430.1$\pm$0.8 [3.7] & 19.5$^*$  &  &\\[5pt]
\end{tabular}
{\footnotesize 
\newline 
Note: $^*$ This is relative to the predicted frequency of the fourth 
harmonic (which is itself of low amplitude and a less reliable 
frequency.\hfill}
\label{tab4}
\end{table}

\begin{figure}
\centerline{\hbox{\psfig{figure=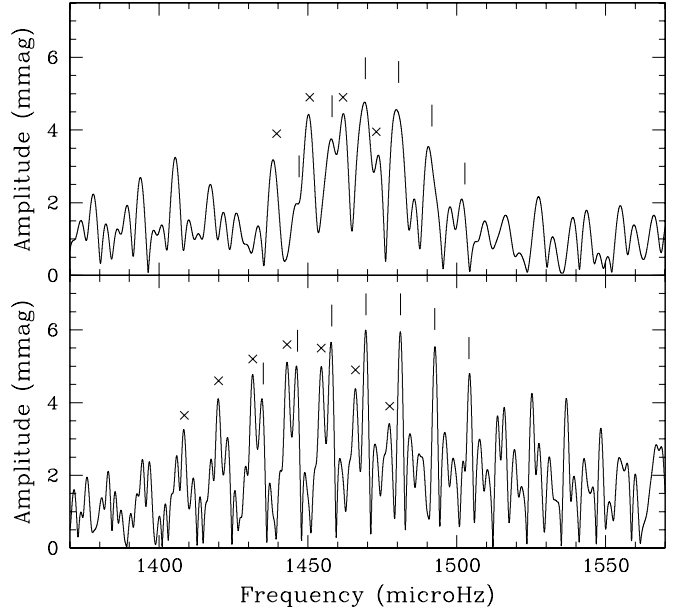,width=8.8cm}}}
  \caption{The Fourier transforms of `2003aw' in a high state (upper panel) and intermediate state (lower panel) centred on
the second harmonic of the superhump frequency. The harmonic (and its aliases) is
marked by the vertical bars, the sideband is marked by crosses.}
 \label{2003awftbu}
\end{figure}

\begin{figure}
\centerline{\hbox{\psfig{figure=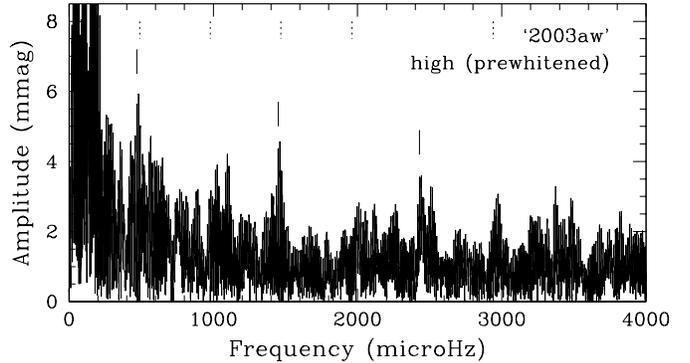,width=8.8cm}}}
  \caption{The Fourier transforms of `2003aw' in a high state after prewhitening 
at the frequencies listed in Table~\ref{tab3} (marked by dotted vertical bars). The solid vertical
bars show the sidebands.}
 \label{2003awftsub}
\end{figure}

In addition to the superhump fundamental and harmonics in the FTs there are signals at neighbouring frequencies. This is illustrated
in Fig.~\ref{2003awftbu} where the second harmonics in both high and intermediate states are accompanied by distinct sidebands of 
comparable amplitudes.  The FT for the high state, prewhitened with the superhump frequency and its harmonics, is given in Fig.~\ref{2003awftsub},
which shows the remaining sideband power at the fundamental, second and fourth harmonics. The sideband frequencies  are listed
in Table~\ref{tab4}. In all cases there is an aliasing problem which could result in the quoted values of $\Delta \nu$ being
decreased by $\sim$ 11.8 $\mu$Hz.

The consistency of the measured values of $\Delta \nu$ in the high state speaks for their reality. The frequency splitting of
19.5 $\mu$Hz is equivalent to a period of 14.2 h, which is suggestively close to the $\sim$ 16 h cycling in brightness described
in Section 3.1. The $\Delta \nu$ measured in the intermediate state is equivalent to 18.5 h.

These results for `2003aw' are in strong contrast to what has been seen in other AM CVn stars. The dwarf nova-like cycling
in CR Boo did not produce sidebands (P1997); whether the cycling in V803 Cen does is not yet known as P2000 found the FT too complicated
to disect. Another contrast is that in AM CVn itself; although constant values of $\Delta \nu$ are seen, they are relative to the
orbital frequency, not the superhump frequency (Skillman et al.~1999). A similar situation obtains in HP Lib (Patterson et al.~2002)
and CR Boo (P1997). There is, as a result, some confusion over what is happening in `2003aw': could we be seeing an orbital hump
rather than a superhump? Is the $\sim$ 16 h cycling due to precession of a disc rather than dwarf nova outbursts?

\section{Discussion}   

Despite its initial discovery via a supernova search, `2003aw' belongs to the class of AM CVn stars and is not a supernova, 
nor is there a galaxy nearby as initially thought (the supposed galaxy on the sky survey plate is in fact a fuzzy image 
of `2003aw' in a low state). Fortunately, however, it was mistaken for a galaxy; `2003aw' might otherwise have remained anonymous.

The short-lived outburst on 2003 March 5, coming when `2003aw' was already in a high state, resembles the behaviour
in some intermediate polars (Schwarz et al.~1988; van Amerongen \& van Paradijs 1989). If these are caused
by temporary increases in $\dot{M}$ from the secondaries then it implies similar behaviour in low mass hydrogen-rich
main sequence stars and very low mass helium white dwarfs. On the other hand, if the short outbursts are caused by
temporary storage and release of mass into the magnetospheres of the intermediate polars (e.g., Taam \& Spruit 1989)
this implies a substantial magnetic field on the primary of `2003aw'.

A pleasing aspect of the new AM CVn star in Hya is that it knows its place in the helium-transferring heirarchy. 
It has previously been pointed out (Warner 1995a,b,c) that the theoretical reduction of $\dot{M}$ with increasing
$P_{orb}$, together with the destablizing effect of irradiation on the secondary (Wu, Wickramasinghe \& Warner 1995)
produces stable high $\dot{M}$ at the shortest periods, VY Scl behaviour at intermediate periods, and low $\dot{M}$, probable
long-interval large amplitude dwarf novae at the longest periods. Each of the recently discovered AM CVn stars, including `2003aw',
fits into this pattern. The current observational boundaries (Table~\ref{tab1}) are:  Stable high $\dot{M}$ for $P_{orb}$(s) $\la 1200$,
VY Scl 1200 $\la P_{orb}$(s) $\la 2500$, Low state $P_{orb}$(s) $\ga 2500$. As more AM CVn stars are discovered, it will be 
interesting to see whether the boundaries are sharp or if there is some fuzziness to them.

\section*{Acknowledgments}
PAW is supported by funds made available from the National Research
Foundation and by strategic funds made available to BW from the
University of Cape Town. BW's research is supported by the University.
We thank Retha Pretorius for kindly taking snapshots of `2003aw'.

\end{document}